\documentclass[review]{elsarticle}
 
\usepackage{lineno,hyperref}
 
\modulolinenumbers[5]

\usepackage[T1]{fontenc}
\usepackage{amsmath}
\usepackage{amssymb}
\usepackage{graphicx}
\usepackage{esint}

 \journal{Journal of Multivariate Analysis}

\biboptions{authoryear}

\begin{document}

\begin{frontmatter}

\title{Entropy measure for the quantification of upper quantile interdependence in multivariate distributions}

\author[myaddress]{Jhan Rodr\'iguez \corref{corrAuthor}}
\cortext[corrAuthor]{Corresponding author}
\ead{Jhan.Rodriguez@iws.uni-stuttgart.de}

\author[myaddress]{Andr\'as B\'ardossy}

\address[myaddress]{Institute for Modelling Hydraulic and Environmental Systems, Universit\"at Stuttgart}

\begin{abstract}
We introduce a new measure of interdependence among the components
of a random vector along the main diagonal of the vector copula, i.e.
along the line $u_{1}=\ldots=u_{J}$, for $\left(u_{1},\ldots,u_{J}\right)\in\left[0,1\right]^{J}$.
Our measure is related to the Shannon entropy of a discrete random
variable, hence we call it an ``entropy index''. This entropy index
is invariant with respect to marginal non-decreasing transformations
and can be used to quantify the intensity of the vector components
association in arbitrary dimensions. We show the applicability of
our entropy index by an example with real data of 5 stock prices of
the DAX index. In case the random vector possesses an extreme value
copula, the index is shown to have as limit the extremal coefficient,
which can be interpreted as the effective number of asymptotically
independent components in the vector. 
\end{abstract}


\end{frontmatter}

\linenumbers

\section*{Introduction\label{sec:Introduction}}

The assessment of the intensity of tail dependence for multivariate
data is an important task in several research areas, such as empirical
finance, econometrics and atmospheric research. Additionally, the
assessment of interdependence among more than two random variables
simultaneously has been indicated as relevant in research fields as
diverse as weather forecasting, empirical finance and spike train
analysis, in neuronal science \cite{bardossy_copula_2009,Cubic1,Dhaene2012357}.

It is useful to have some graphical tool to visualize the intensity
of the association or dependence as one approaches the tail of the
distribution. For example, the Chi-plot (\cite{fisher2001graphical})
has been used to pin down specific characteristics of tail behavior
by \cite{abberger2005simple}.

We present in this paper an index of association along the main diagonal
of the copula of the distribution, i.e. along the line $u_{1}=\ldots=u_{J}$,
for $\left(u_{1},\ldots,u_{J}\right)\in\left[0,1\right]^{J}$. This
index can be plotted to check for interdependence intensity at the
uppermost quantiles of the distribution, as in \cite{abberger2005simple},
but can also be readily applied to a random vector of dimension greater
than two. 

The rest of this paper is organized as follows: In section \ref{sec:The-entropy-index}
we introduce the new association measure, to which we refer as an
``entropy index''. In section \ref{sec:Examples} we apply the index
to explore the type of dependence for 2, 3 and 4 dimensional marginal
distributions of a real data set, and show how it can be used to evaluate
goodness of tail fit for three different models fitted to the data.
In section \ref{sec:Relation-with-extremal} we exhibit the relation
of our entropy index with the extremal coefficient (see, for example,
chapter 8 of \cite{beirlant2004statistics}, and \cite{schlather2003dependence}),
if the distribution of the analyzed vector is in the domain of attraction
of an extreme value distribution. We end the paper with some conclusions
and further interesting explorations of the entropy index.

\section{The entropy index\label{sec:The-entropy-index}}

We begin by recapitulating the \textquotedbl{}congregation measure\textquotedbl{}
used by \cite{bardossy_copula_2009} and \cite{bardossy_multiscale_2012},
for the sake of model validation. A modification of this measure constitutes
the interdependence measure we introduce in this paper. 

Let $\mathbf{X}=\left(X_{1},\ldots,X_{J}\right)$ be a random vector
with copula $C$, so that 
\[
F_{\mathbf{X}}\left(X_{1},\ldots,X_{J}\right)=C\left(F_{1}\left(X_{1}\right),\ldots,F_{J}\left(X_{J}\right)\right)
\]
 where $F_{\mathbf{X}}$ is the probability distribution function
of $\mathbf{X}$ and $F_{1},\ldots,F_{J}$ its marginal distribution
functions. Our analysis focuses on the standardized random vectors
\[
\mathbf{U}=\left(U_{1},\ldots,U_{J}\right)=\left(F_{1}\left(X_{1}\right),\ldots,F_{J}\left(X_{J}\right)\right)
\]

Set a threshold percentile, $b\in\left(0,1\right)$. Select a set
of indexes $\left(j_{i_{1}},\ldots j_{i_{K}}\right)$, with $1\leq j_{i_{1}}<\ldots<j_{i_{K}}\leq J$.
For the analysis of the components of $\mathbf{U}$, define binary
random variables 
\begin{equation}
\varsigma_{b}\left(j_{i_{k}}\right)=\begin{cases}
1, & U_{j_{k}}>b\\
0, & U_{j_{i_{k}}}\leq b
\end{cases}
\end{equation}

This results in a discrete random vector, $\varsigma_{b}=\left(\varsigma_{b}\left(j_{i_{1}}\right),\ldots,\varsigma_{b}\left(j_{i_{K}}\right)\right)$.
The congregation measure introduced by \cite{bardossy_copula_2009}
is defined to be the entropy of a sub-vector of $\varsigma$,
\begin{multline}
H_{b}\left(U_{j_{i_{1}}},\ldots,U_{j_{i_{K}}}\right)=\\
-\sum_{j_{i_{1}},\ldots,j_{i_{K}}}\Pr\left(\varsigma_{b}\left(j_{i_{1}}\right),\ldots,\varsigma_{b}\left(j_{i_{K}}\right)\right)\log\left(\Pr\left(\varsigma_{b}\left(j_{i_{1}}\right),\ldots,\varsigma_{b}\left(j_{i_{K}}\right)\right)\right)\label{eq:entropy_measure}
\end{multline}

That is, the measure is defined as the (Shannon) entropy of the joint
distribution of the binary variables just defined. A higher value
of this measure indicates less association, and vice versa.

Note that if the copula $C$ is the independence copula, 
\[
C\left(u_{1},\ldots,u_{J}\right)=C^{\sim}\left(u_{1},\ldots,u_{J}\right)=\prod_{j=1}^{j=J}u_{j}
\]
then the measure is constantly
\[
H_{b}=-J\left(b\log\left(b\right)+\left(1-b\right)\log\left(1-b\right)\right)
\]
whereas if $C$ is the co-monotonic copula, 
\[
C\left(u_{1},\ldots,u_{J}\right)=\min\left(u_{1},\ldots,u_{J}\right)
\]
then the measure is also constant:
\[
H_{b}=-\left(b\log\left(b\right)+\left(1-b\right)\log\left(1-b\right)\right)
\]

Between these two extremes lies the congregation measure, upon application
to any given copula. Our entropy index is given by
\begin{equation}
S{}_{b}\left(\mathbf{U}\right)=\frac{H_{b}\left(U_{j_{i_{1}}},\ldots,U_{j_{i_{K}}}\right)}{-\left(b\log\left(b\right)+\left(1-b\right)\log\left(1-b\right)\right)}\label{eq:Original_index}
\end{equation}

This index quantifies the deviance from the totally dependent case.
It is 1 in case of total dependence, and $J$ in case of independence
among the components of $\mathbf{U}$. Evidently, it can be used regardless
of the dimension of $\mathbf{U}$, while keeping its interpretability
as quantification of deviance from total independence. 

In the following, we obviate the dependence on $\mathbf{U}$ in order
to make notation simpler.

An alternative, more general definition of the index, uses the so-called
Tsallis entropy instead of the standard Shannon entropy. The Tsallis
entropy includes an additional parameter $\alpha\in\left(0,+\infty\right)$
and is defined by 
\[
H_{b}^{\alpha}=\frac{1}{\alpha-1}\left(1-\sum_{j_{i_{1}},\ldots,j_{i_{K}}}\Pr\left(\varsigma_{b}\left(j_{i_{1}}\right),\ldots,\varsigma_{b}\left(j_{i_{K}}\right)\right)^{\alpha}\right)
\]
which reduces to $H_{b}$ by letting $\alpha\rightarrow1$. In this
paper, we use the Tsallis entropy definition only as a technical tool
in   \ref{sec:Proof-of-convergence}, for proving a convergence
result.

\section{Example of applicability\label{sec:Examples}}

We consider the stock prices of four components of the German DAX
index, namely ADIDAS, ALLIANZ, BASF and BAYER, re-labeled in the following
as components 1,2,3 and 4, respectively. The daily data spans the
period from January 3th 2000 through June 30th 2014. The data used
is available at www.finanzen.de.

For each stock $j=1,2,3,4$, we shall not consider the closing stock
price at day $t$, $p_{t,j}$, but rather the log-returns 
\begin{equation}
r_{t,j}=100\times\left(\log\left(p_{t,j}\right)-\log\left(p_{t-1,j}\right)\right)
\end{equation}

This results in 4 time series, one for each stock. We use for our
analysis of this 4-dimensional data set a model similar to that of
\cite{abberger2005simple} and \cite{dias2004dynamic}. The approach
consists in fitting a time series model to each of the time series,
independently, and then fitting a multivariate distribution to the
resulting (presumably) iid vector formed out of the residuals of the
time series. 

To each of the four log-returns time series, we fit a GARCH(1,1) model.
We obviate the $j$ index in order to simplify notation, but the first
equation below should read $r_{t,j}=\mu_{j}+a_{t,j}$, and so on.
The model for each of the four time series is

\begin{eqnarray}
r_{t} & = & \mu+a_{t}\\
a_{t} & = & \sigma_{t}\times\epsilon_{t}\\
\sigma_{t}^{2} & = & \alpha_{0}+\alpha_{1}a_{t-1}^{2}+\beta_{1}\sigma_{t-1}^{2}
\end{eqnarray}

Under the assumption that the $\epsilon_{t}$ are iid with $E\left(\epsilon_{t}\right)=0$
and $Var\left(\epsilon_{t}\right)=1$. Two typical assumptions for
the so-called ``standardized shocks'', $\epsilon_{t}$, are $\epsilon_{t}\sim N\left(0,1\right)$
and $\sqrt{\frac{\nu}{\left(\nu-2\right)}}\times\epsilon_{t}\sim t_{\nu}$,
for $\nu>2$. For details, the reader is referred, for example, to
\cite{tsay2005analysis}. The idea of the GARCH model is to reproduce
the clustering in variance, not explainable by linear time series
models (like the ARMA model), often present in financial time series.

A GARCH(1,1) model was fitted independently to each log-returns time
series using the garchFit function of package fGarch of the R statistical
software (\cite{RCran}). Estimation was performed using the Quasi
maximum likelihood (QMLE) option, which is robust against miss-specification
of the standardized shocks distribution. 

The vectors of standardized shocks $\epsilon_{t}=\left(\epsilon_{t,1},\ldots,\epsilon_{t,4}\right)$,
$t=1,...,3686$ become now our object of study, or ``observed data''.
They are assumed to be (sufficiently) temporally independent, but
there can be contemporaneous interdependence, with which we now deal.
This contemporaneous interdependence, analogous to (\cite{abberger2005simple,dias2004dynamic})
can be modeled by a copula model. Two models used in the literature
are the Gaussian copula and the Student copula with unknown (i.e.
to estimate) degrees of freedom, $\nu$. Their correlation matrices
are also estimated in the process.

We fit in the following three models to the copula of random vector
$\epsilon$. The first two are a Gaussian copula, and a Student copula.
These copulas are fitted to the transformed standardized shocks, $u_{t,j}$,
$j=1,\ldots,4$, $t=1,\ldots,3686$, obtained by
\begin{equation}
u_{t,j}=F_{n,j}\left(\epsilon_{t,j}\right)
\end{equation}
where $F_{n,j}$ stands for the empirical distribution function of
the respective component, $j$, and is given by
\[
F_{n,j}\left(\epsilon_{t,j}\right)=\frac{\#\left\{ \epsilon_{s,j}:\epsilon_{s,j}\leq\epsilon_{t,j}\right\} }{3687}
\]

The third model explored here is not a copula model. We represent
the density of $\epsilon$ by a mixture of five multivariate normal
distributions 
\[
f\left(\epsilon\right)=\sum_{k=1}^{5}w_{k}g_{k}\left(\epsilon\right)
\]
where each Normal distribution is allowed to have its own mean vector
and covariance matrix; such mean vectors, covariance matrices and
the weights $\left(w_{1},\ldots,w_{k}\right)$ are estimated on the
basis of the available data.

We used the R software to fit all the respective model parameters:
Function fitCopula of package \emph{copula} to fit the copula models
parameters; and function init.EM of package \emph{EMcluster} to fit
the mixture model, which estimates the mixture parameters by the Expectation
Maximization algorithm. The fitted degrees of freedom for the Student
copula model were 7.50, which are those of a model with non-negligible
tail dependence.

We proceed now to the analysis of the joint association of 2,3 and
4 dimensional joint marginals of vector $\epsilon$, both of the observed
data and of data simulated from the three fitted models. To this end,
we use the first 2, 3 and 4 components of $\epsilon$, respectively.
Our analysis consists in computing the entropy index defined in section
\ref{sec:The-entropy-index} for increasing quantile thresholds, $b\in\left(0.85,1.00\right)$,
at the upper part of the distribution. Specifically, we use threshold
values $b=.850,.855,\ldots,.995$.

We shall see how the degree of association of each of the models along
the line $u_{1}=\ldots=u_{4}=b$ is, as compared to that of the observed
data. This gives us an idea of the adequacy the modeled interdependence,
for the 2,3 and 4-dimensional marginals, as one focuses on the uppermost
part of the distribution. 

\begin{figure}
\begin{centering}
\includegraphics[width=0.33\textwidth]{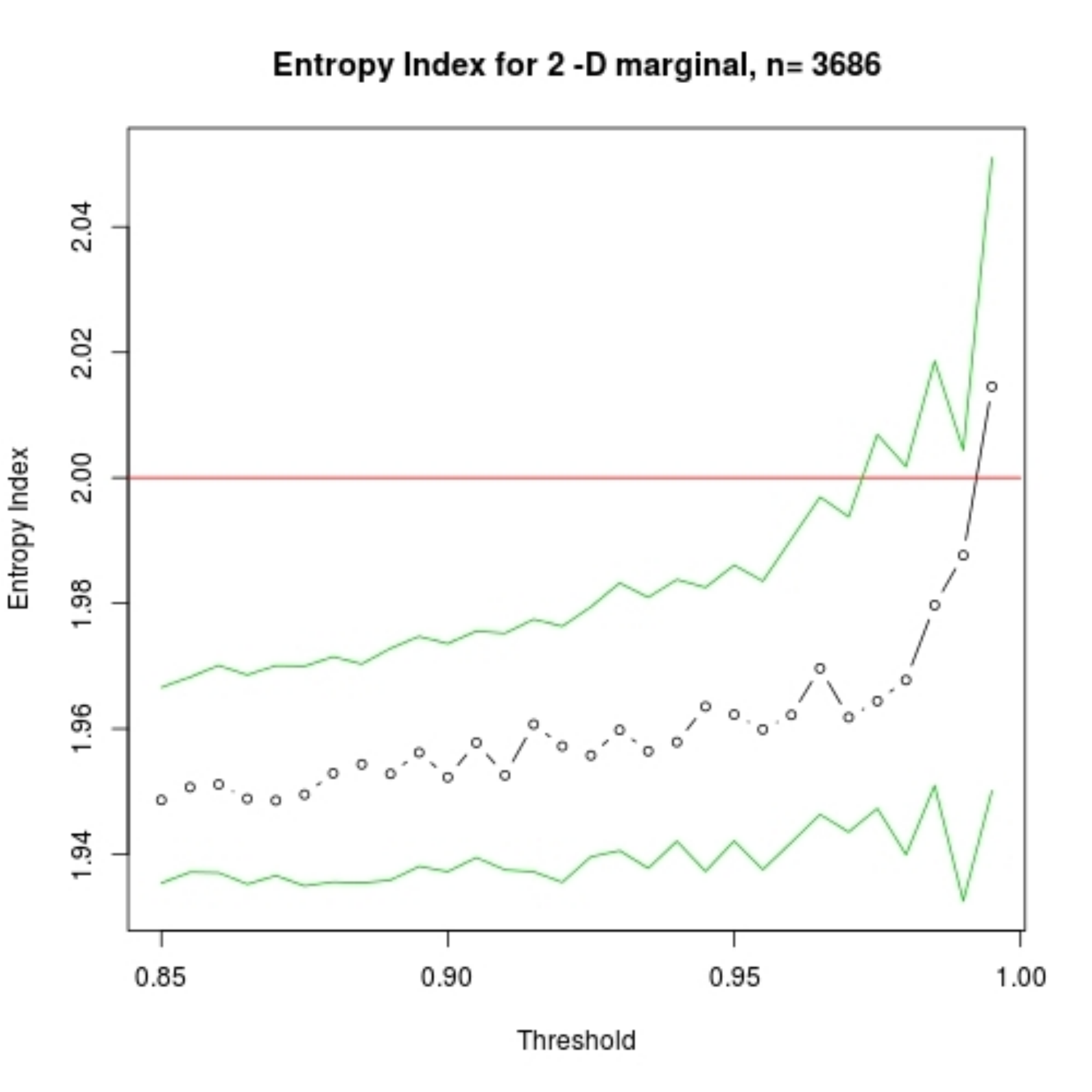}\includegraphics[width=0.33\textwidth]{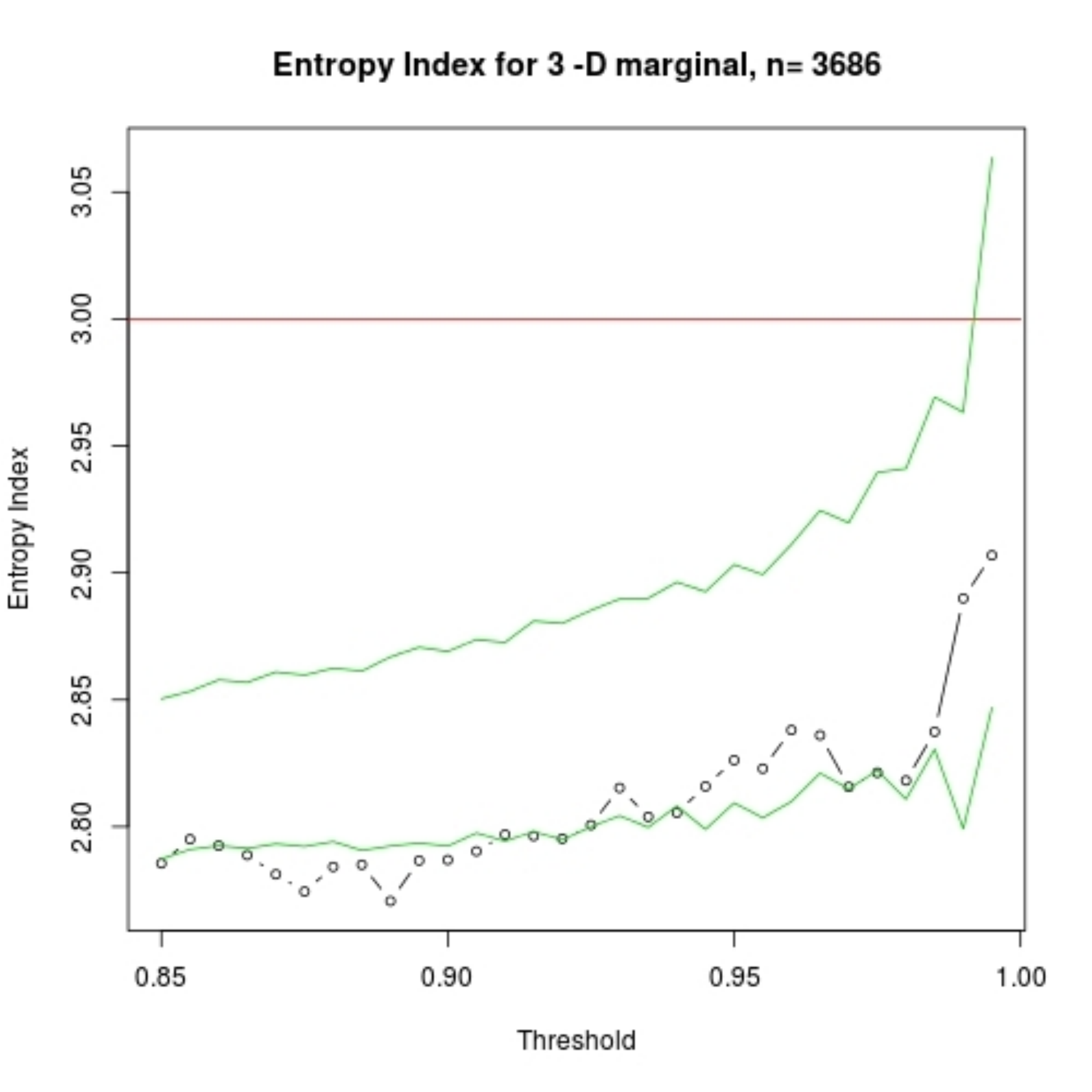}\includegraphics[width=0.33\textwidth]{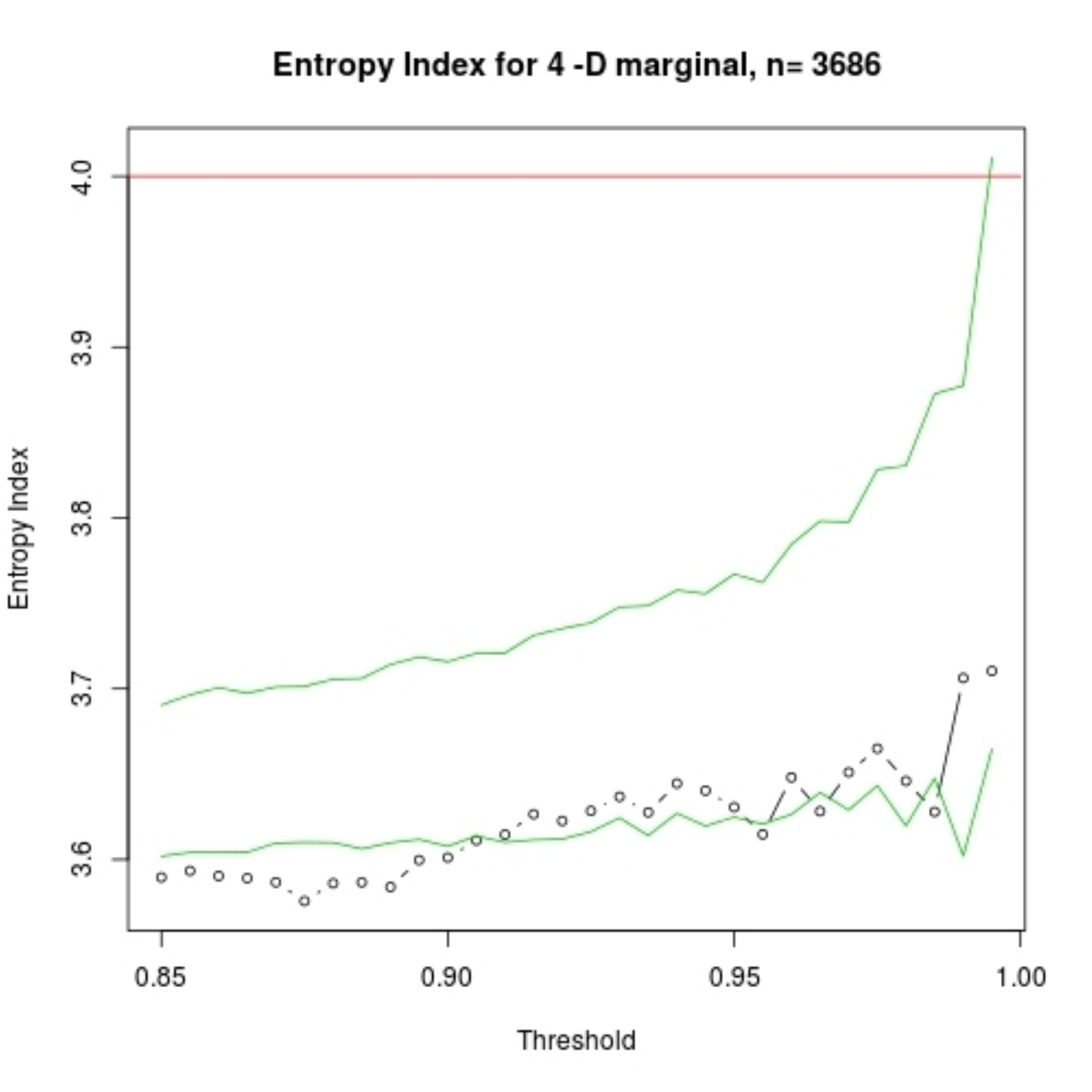}
\par\end{centering}

\centering{}\protect\caption{\label{fig:Entropy-index-normal}Entropy index computed for selected
quantile thresholds: Data (Black dotted line) and Monte Carlo confidence
interval for data coming from the fitted Gaussian copula.}
\end{figure}

In figure \ref{fig:Entropy-index-normal} we show the entropy index
computed for the observed data at the indicated thresholds, $b$,
given by the black lines and points. The green lines added correspond
to a 95\% confidence interval for data obtained from the fitted Gaussian
copula model. The confidence interval is based on the generation of
500 data sets, each of size 3686, of the fitted Gaussian copula model,
and the computation of the entropy index for the thresholds $b=.850,.855,\ldots,.995$,
as had been done for observed data.

From figure \ref{fig:Entropy-index-normal}, we see that the type
of association in the observed data, as represented by the entropy
index, is similar in the two-dimensional marginals to that of the
Gaussian copula. Data association is however systematically stronger,
for the 3 and 4 dimensional distributions considered: the black line
sticks to the bottom of the confidence interval, sometimes even stepping
out of it. 

This is another warning about the need to validate multivariate statistical
models by statistics that consider more than two components at a time,
when joint interaction among more than two components is relevant
for the problem at hand (cf. \cite{bardossy_copula_2009,bardossy_multiscale_2012,2014arXiv1406.2501R}). 

\begin{figure}
\begin{centering}
\includegraphics[width=0.33\textwidth]{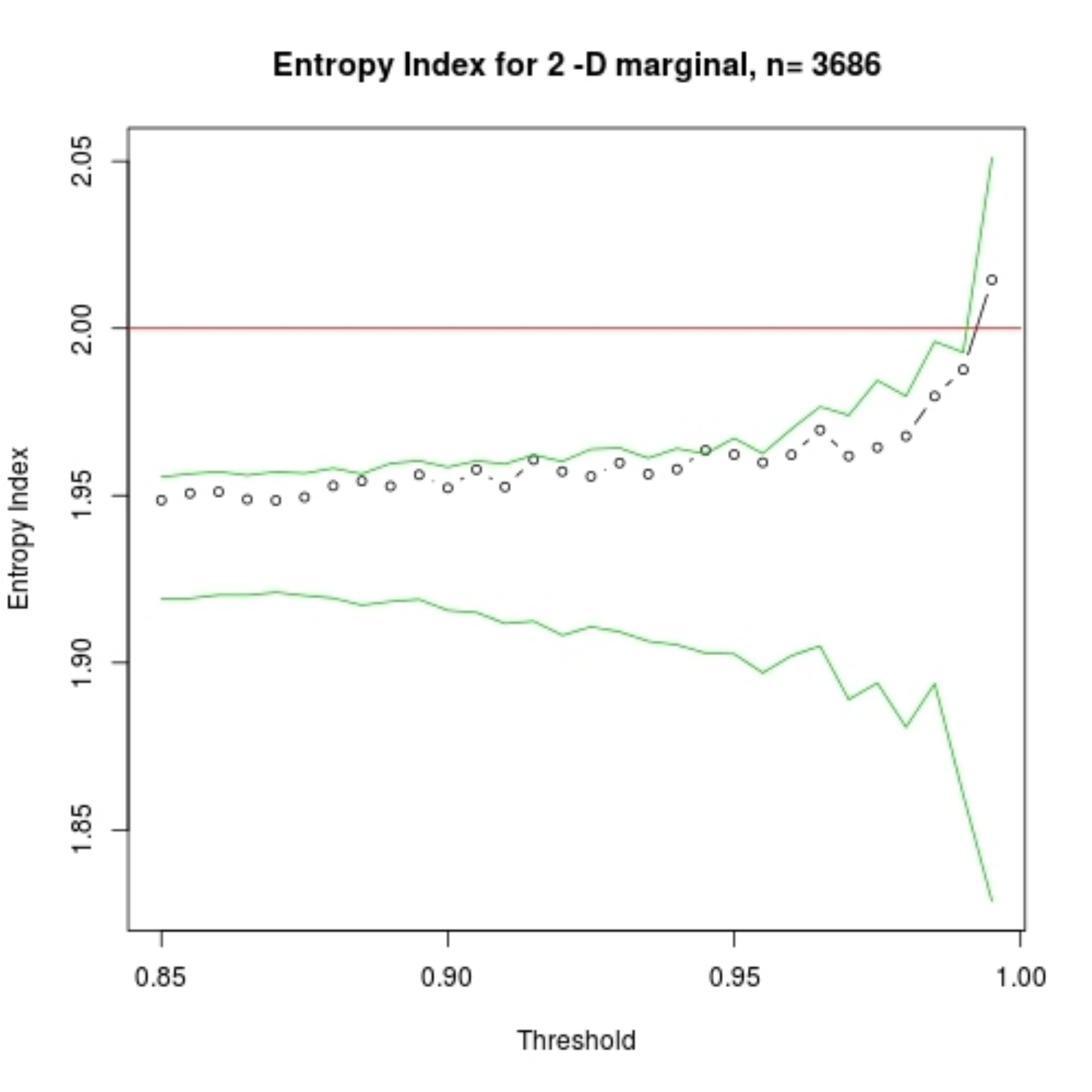}\includegraphics[width=0.33\textwidth]{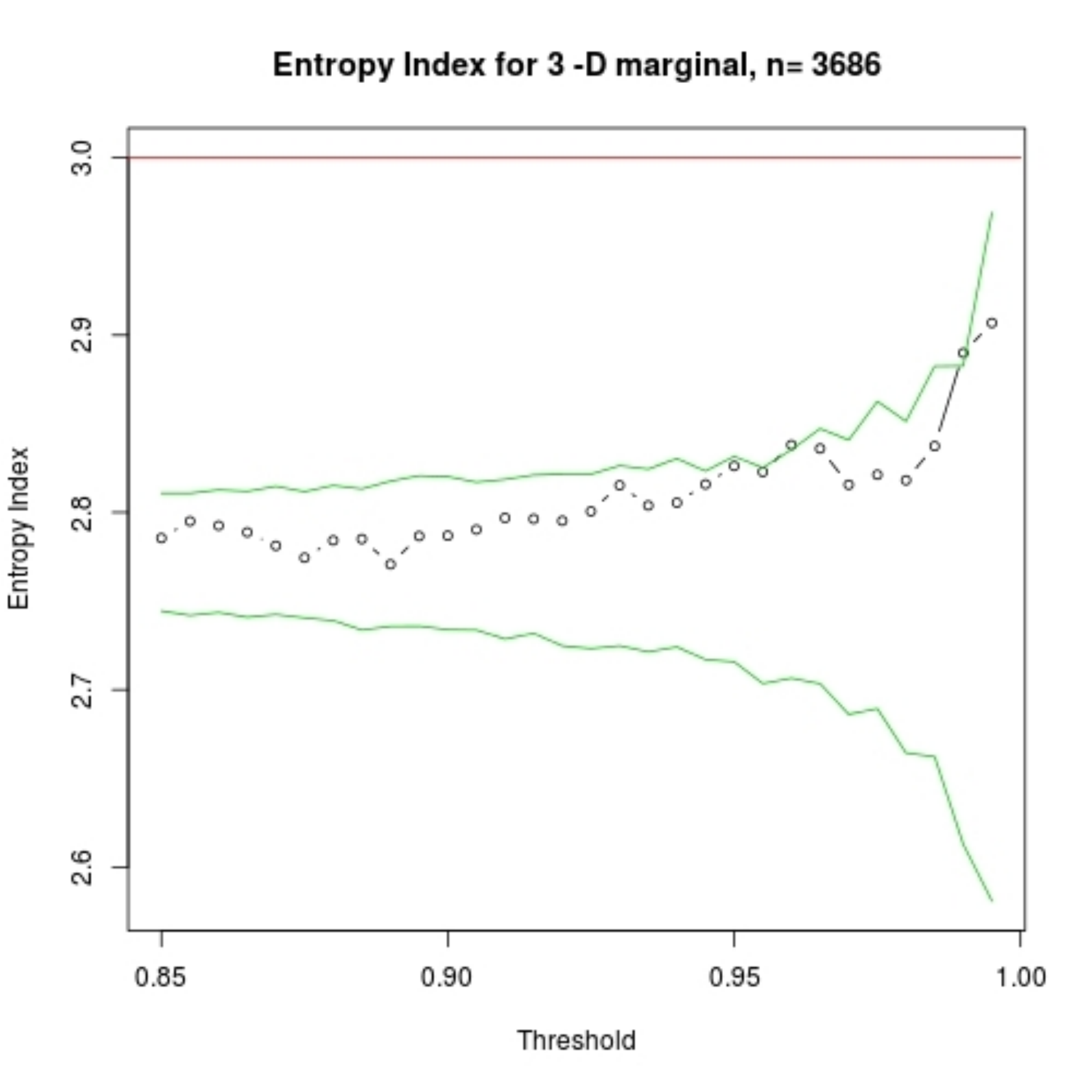}\includegraphics[width=0.33\textwidth]{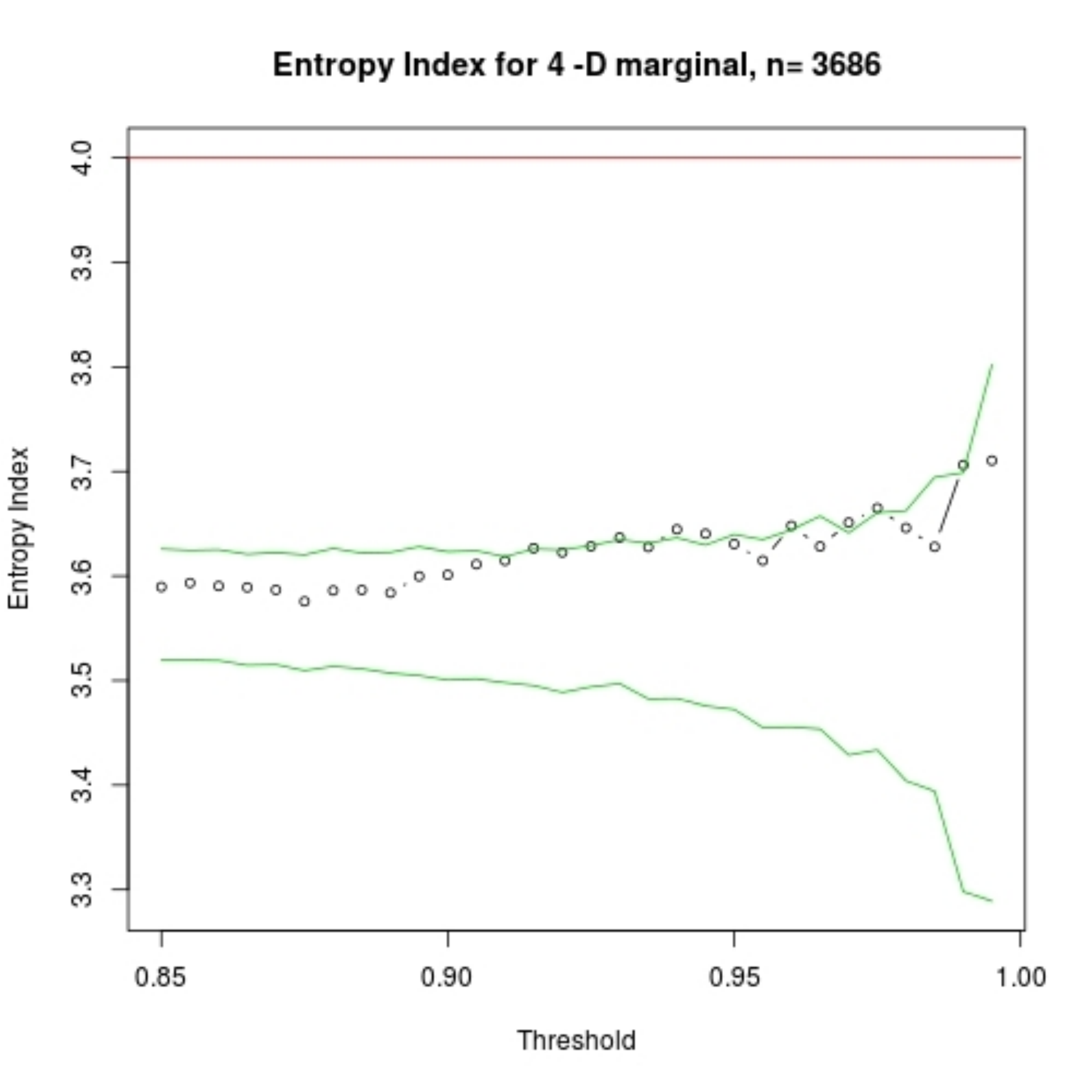}
\par\end{centering}

\centering{}\protect\caption{\label{fig:Entropy-index-student}Entropy index computed for selected
quantile thresholds: Data (Black dotted line) and Monte Carlo confidence
interval for data coming from the fitted Student-t copula.}
\end{figure}

In figure \ref{fig:Entropy-index-student}, we show the same type
of plot as before, but for the student copula model. We note that
the interdependence is somewhat exaggerated, even for the 2-dimensional
marginal considered. This exaggeration is clearer for the 4-dimensional
joint distribution. Even if the asymptotic tail dependence were right,
the representation of the interdependence among the process variables
is not adequate for high (though not extreme) quantiles of the joint
distributions shown. 

\begin{figure}
\centering{}\includegraphics[width=0.33\textwidth]{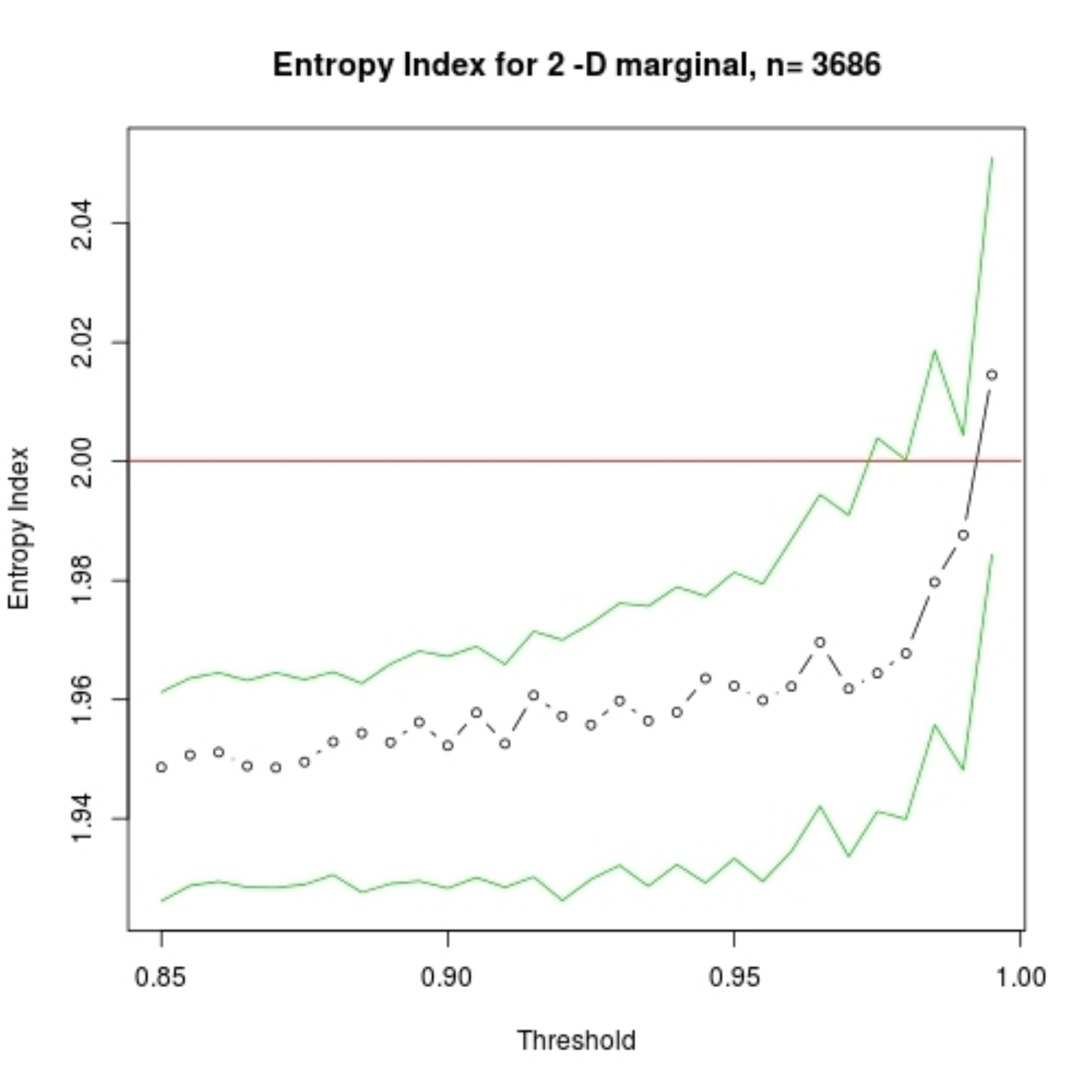}\includegraphics[width=0.33\textwidth]{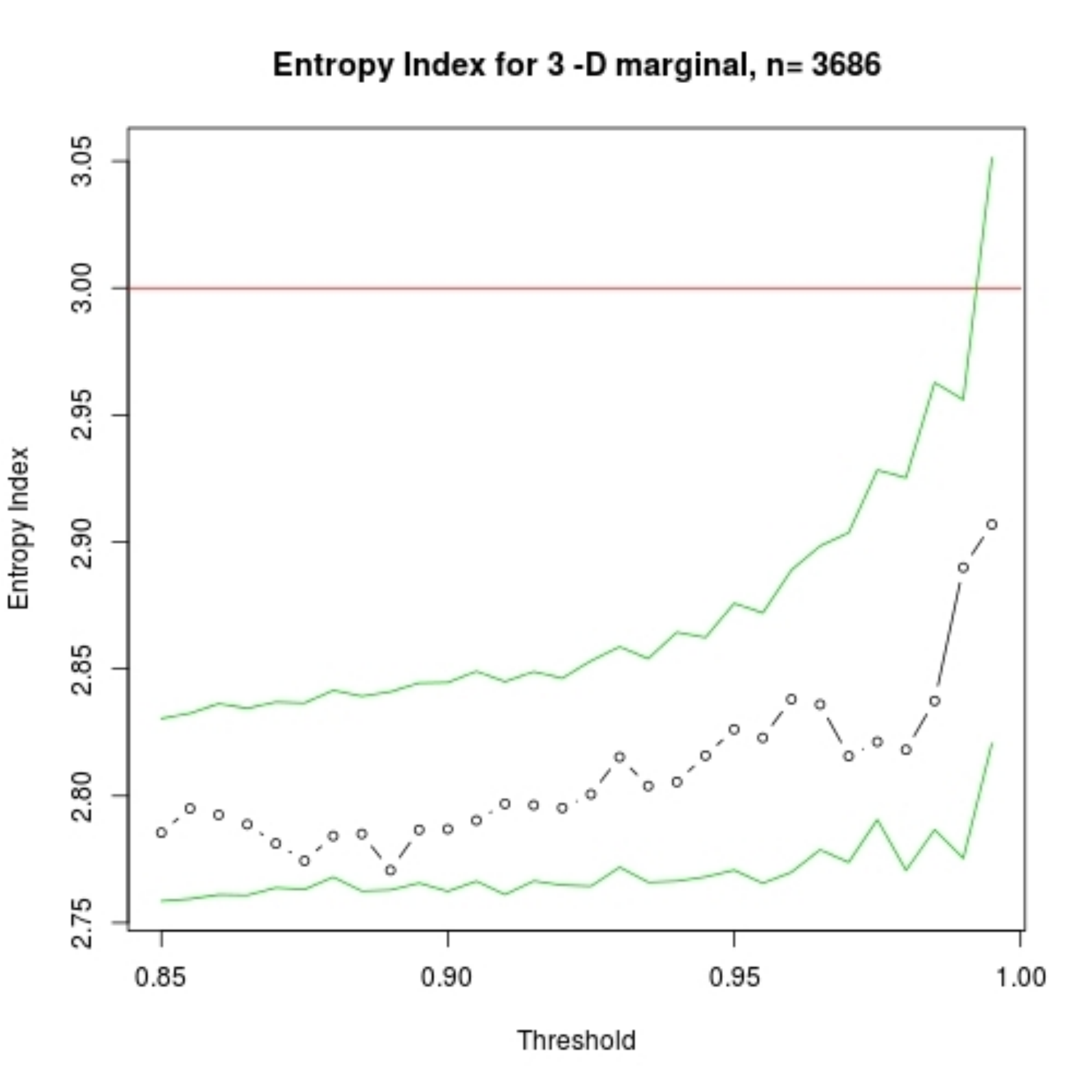}\includegraphics[width=0.33\textwidth]{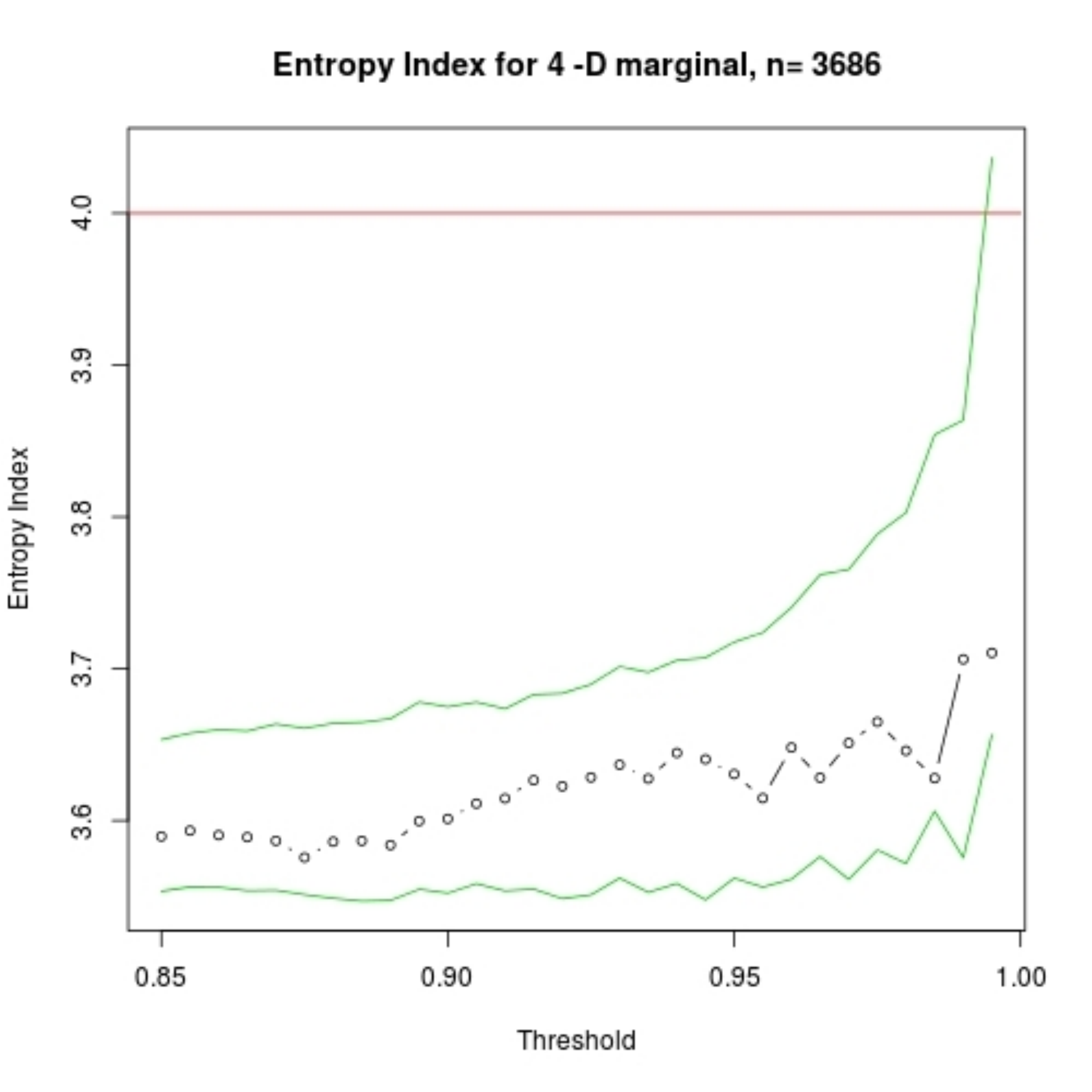}\protect\caption{\label{fig:Entropy-index-mixture}Entropy index computed for selected
quantile thresholds: Data (Black dotted line) and Monte Carlo confidence
interval for data coming from the fitted mixture of Normal distributions.}
\end{figure}

The same procedure as above is repeated with the mixture model. The
result is shown in figure \ref{fig:Entropy-index-mixture}. Note that
the black line, corresponding to the measure of interdependence for
data, lies roughly at the center of the confidence interval, for each
of the joint 2, 3 and 4-dimensional marginals presented. 

This is an indication that, in terms of the type of association measured
by our entropy index, the mixture model is a more realistic representation
of the process, as compared with the other two models. That it is
a better representation of the variables interdependence at the section
of the distribution just before the extreme value region, as one \emph{approaches}
that region. 

This better fit is not much of a surprise, since 5 mixture components
provide considerable modeling flexibility. Our point here is not to
favor the a specific, over-parameterized model, but to show how one
can notice important deficiencies in the fit of a given model in a
$d\geq2$ -dimensional setting.

Sometimes finance data, like the one here shown, is subject to tail
dependence (cf. \cite{abberger2005simple}). In that case, an even
better model would be a mixture of Student distributions, with degrees
of freedom higher than 7, each. In this way, the joint association
for high quantiles would not be exaggerated, while the asymptotic
tail dependence will not be zero, as in the case of a mixture of multivariate
normal distributions.

\section{Relation with extremal coefficient\label{sec:Relation-with-extremal}}

An extreme value copula is the copula of an extreme value distribution,
$G$, and can be characterized by the following stability condition:
A copula on $\left[0,1\right]^{J}$ is of the extreme value type if,
and only if, 
\begin{equation}
C_{G}^{s}\left(u_{1},\ldots,u_{J}\right)=C_{G}\left(u_{1}^{s},\ldots,u_{J}^{s}\right)\label{eq:EV_copula}
\end{equation}
for all $s>0$.

For $b\in\left[0,1\right]$, extreme value copulas fulfill the relation
\begin{equation}
C_{G}\left(b,\ldots,b\right)=b^{\theta}\label{eq:extremal_coeff_EVCopula}
\end{equation}
for some $\theta\in\left[1,J\right]$. This parameter $\theta$ receives
the name of \emph{extremal coefficient }(see, for example chapter
8 of \cite{beirlant2004statistics}). It can be thought of as the
asymptotic ``effective number of independent variables'' (\cite{schlather2003dependence})
of $\mathbf{X}$. Its value lies between 1 and $J$, in case of asymptotic
perfect dependence and complete independence, respectively. 

Assume that the distribution of $\mathbf{X}=\left(X_{1},\ldots,X_{J}\right)$
is in the domain of attraction of an extreme value distribution, $G$,
i.e. $F_{\mathbf{X}}\in D\left(G\right)$. Since (\cite{beirlant2004statistics},
p. 282) 
\begin{equation}
b\rightarrow1^{-}\Rightarrow C_{F}\left(b,\ldots,b\right)\rightarrow C_{G}\left(b,\ldots,b\right)\label{eq:similarity_upper_quantile}
\end{equation}
one then has that:
\begin{equation}
C_{F_{\mathbf{X}}}\left(b,\ldots,b\right)\rightarrow b^{\theta}\label{eq:extremal_coeff}
\end{equation}
so that, also for the copula of $\mathbf{X}$, interdependence along
the main diagonal of the copula is determined by the extremal coefficient.
The extremal coefficient is the same in both cases.

The limit of our entropy index is precisely the extremal coefficient
of the copula of $\mathbf{\mathbf{X}}$. Namely, 
\begin{equation}
\lim_{b\rightarrow1^{-}}S_{b}\left(\mathbf{U}\right)=\theta\label{eq:Limit_convergence}
\end{equation}
for $\mathbf{U}=\left(U_{1},\dots,U_{J}\right)$, $U_{j}=F_{j}\left(X_{j}\right)$,
and $F_{j}$ the marginal distribution of $X_{j}$, for $j=1,\ldots,J$. 

The proof of (\ref{eq:Limit_convergence}) is rather technical, so
it is relegated to the appendix.

The following is thus an application of the entropy index: For a random
vector $\mathbf{X}$ whose distribution is in the domain of attraction
of an extreme value distribution, we can explore how ``fast'' its
asymptotic effective number of independent components is approached. 

\begin{figure}
\begin{centering}
\includegraphics[width=0.75\textwidth]{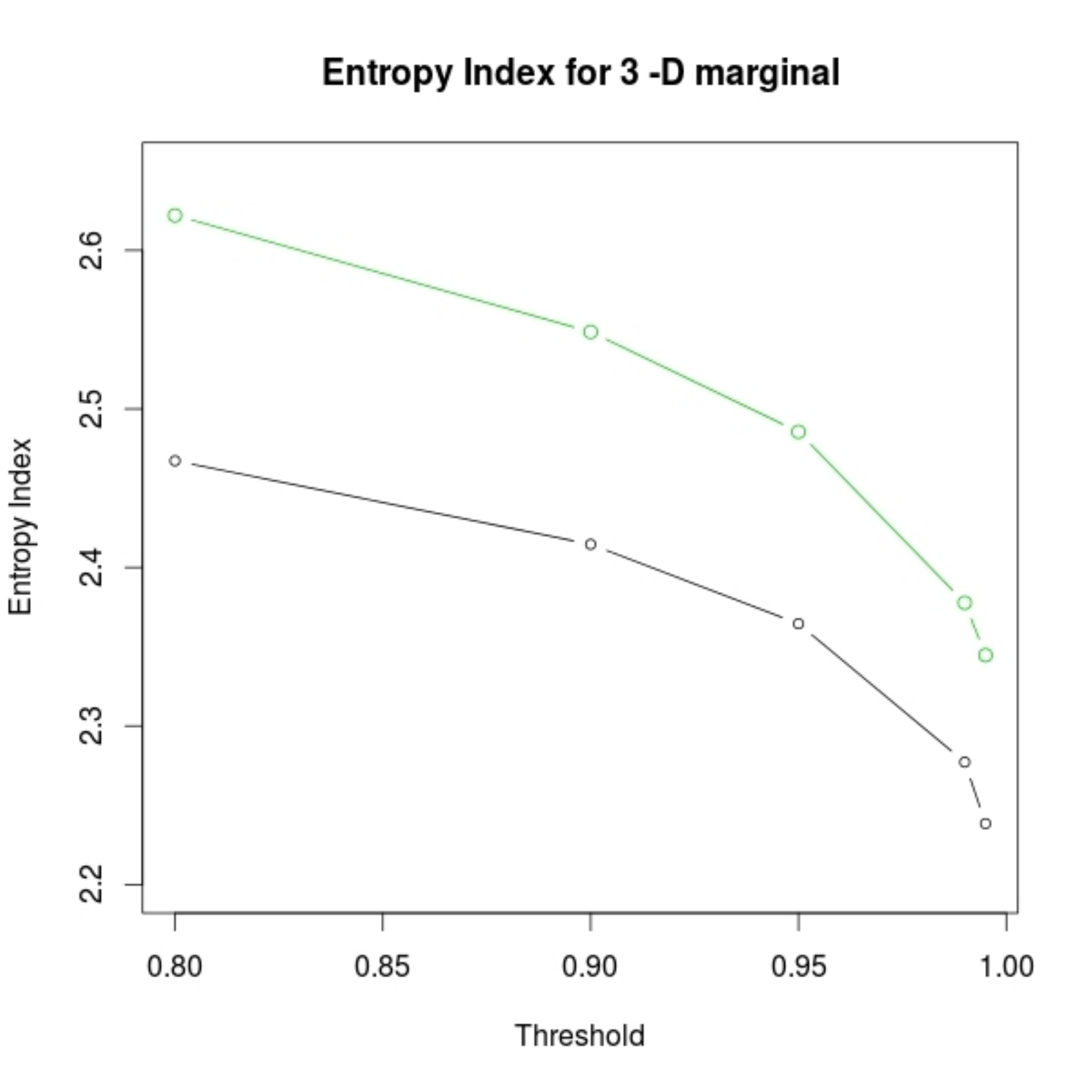}\protect\caption{\label{fig:Comparison-of-entropy-Gumbel-Student}Comparison of entropy
index for the 3-dimensional Gumbel copula (green) and Student copula
(black). Thresholds are $b=.8,.9,.95,.99,.995$. Both distributions
have extremal coefficient $\theta=2$, towards which they approach.
However, note that the Student copula approaches its asymptotic number
of effectively independent components faster.}

\par\end{centering}

\end{figure}

As an example, see figure \ref{fig:Comparison-of-entropy-Gumbel-Student},
where the entropy index is computed for two different copulas, for
thresholds $b=.8,.9,.95,.99,.995$. Figure \ref{fig:Comparison-of-entropy-Gumbel-Student}
is the result of computing the entropy index for the mentioned thresholds
to a simulated sample of size $n=10^{6}$, for each distribution;
so figure \ref{fig:Comparison-of-entropy-Gumbel-Student} is a good
approximation to a figure based on exact, analytic computations.

The green line corresponds to a Gumbel 3-dimensional copula, 
\begin{equation}
C\left(u_{1},u_{2},u_{3}\right)=\exp\left\{ -\left[\left(-\log\left(u_{1}\right)\right)^{\xi}+\ldots+\left(-\log\left(u_{3}\right)\right)^{\xi}\right]^{\frac{1}{\xi}}\right\} \label{eq:gumbel_copula}
\end{equation}
for which the extremal coefficient is $\theta=3^{\frac{1}{\xi}}$. 

The black line at figure \ref{fig:Comparison-of-entropy-Gumbel-Student}
corresponds to a student copula with $\nu$ degrees of freedom and
correlation matrix $\rho\in\mathbb{R}^{3\times3}$, for which the
extremal coefficient is given by 
\begin{equation}
\theta=\sum_{j=1}^{3}T_{2,\nu+1,\rho_{-j,-j}}\left(\frac{\sqrt{\nu+1}}{\sqrt{1-\rho_{i,j}}}\left(1-\rho_{i,j}\right),i\neq j\right)\label{eq:extremal_student}
\end{equation}
where $T_{2,\nu+1,\rho_{-j,-j}}$ stands for the 2-dimensional Student
cumulative distribution function with $\nu+1$ degrees of freedom
and dispersion matrix $\rho_{-j,-j}$. In turn, $\rho_{-j,-j}$ is
the $2\times2$ matrix resulting from removing row and column $j$
from $\rho$. That (\ref{eq:extremal_student}) contains the extremal
coefficient in question can be readily seen from Theorem 2.3, equation
2.8, of \cite{nikoloulopoulos2009extreme}.

Parameters $\xi$, $\nu$ and $\rho$ were selected in such a way
that, for both distributions, the extremal coefficient is $\theta=2$.
The specific parameters used can be found in the appendix. 

In spite of having the same asymptotic ``effective number of independent
components'', the association among the components of the Student
copula is systematically stronger (in terms of the entropy index)
before reaching that limit. 

This is an additional support in favor of our entropy index as a means
of analyzing interdependence carefully at the uppermost part of the
distribution. This detailed analysis can be useful for goodness of
fit purposes.

\section{Conclusion and future work\label{sec:Conclussion-and-future}}

We have introduced a tool that is useful for exploring the intensity
of association at the joint upper quantiles of a multivariate distribution.
This tool is not limited to 2-dimensional distributions. We can have
a measure of how strong or weak is the association just before the
extreme value case, where the intensity of dependence may also be
important for some applications.

The limit of the entropy index, in case of a vector having a distribution
in the domain of attraction of an extreme value distribution, is the
extremal coefficient, $\theta$. This coefficient can be interpreted
as the asymptotic effective number of independent components in the
random vector. So, using the entropy index presented in this article,
we can have an idea of how fast this asymptotic value is reached by
the vector components.

\subsection*{Acknowledgments}
This paper stems from the Ph.D research of the first author, which
was funded by a scholarship of the German Academic Exchange Service
(DAAD). This Ph.D work was carried out within the framework of the
ENWAT program at the University of Stuttgart.

\section*{References}
\bibliography{Report_QE}

\appendix

\section{Proof of convergence to Extremal coefficient\label{sec:Proof-of-convergence}}

We present here the proof of equation (\ref{eq:Limit_convergence}). 

We assume in the following that random vector $\mathbf{U}\in\left[0,1\right]^{J}$
has an extreme value copula, so that $C\left(b,\ldots,b\right)=b^{\theta}$.
By virtue of equation (\ref{eq:similarity_upper_quantile}), the argument
could be repeated for $\mathbf{U}\in\left[b_{0},1\right]^{J}$, with
$b_{0}$ sufficiently close to 1, discarding the hypothesis of an
extreme value copula, but assuming the original random vector in the
domain of attraction of an extreme value distribution. 

Let random vector $\mathbf{U}\in\left[0,1\right]^{J}$ have a copula,
$C$, of the extreme value type. We show in this section that, for
any $\alpha>1$, 
\begin{equation}
\lim_{b\rightarrow1^{-}}T_{b}^{\alpha}\left(\mathbf{U}\right)=\theta\label{eq:limite_index}
\end{equation}

Then, by letting $\alpha\rightarrow1^{+}$, we shall have that $\lim_{b\rightarrow1^{-}}S_{b}\left(\mathbf{U}\right)=\theta$. 

To simplify notation, dependence of $S_{b}$ on random vector $\mathbf{U}$
is taken for granted in the following, so that, for example, $S_{b}:=S_{b}\left(\mathbf{U}\right)$,
and so on.

To prove equation (\ref{eq:limite_index}) we shall provide ``sandwich''
functions $g_{1}\left(b\right)$ and $g_{2}\left(b\right)$, such
that $g_{1}\left(b\right)\leq T_{b}^{\alpha}\leq g_{2}\left(b\right)$
for all $b$, $b_{0}<b<1$ (for some sufficiently large $b_{0}$).
Of these auxiliary functions, it will be easy to show that 
\[
\lim_{b\rightarrow1^{-}}g_{1}\left(b\right)=\lim_{b\rightarrow1^{-}}g_{2}\left(b\right)=\theta
\]
whence, necessarily, one must have (\ref{eq:limite_index}). 

Of all the probabilities appearing at (\ref{eq:entropy_measure}),
the most straightforward to identify is 
\[
\Pr\left(\varsigma_{b}\left(1\right)=0,\ldots,\varsigma_{b}\left(J\right)=0\right)=C\left(b,\ldots,b\right)
\]
if we have the function defining $C$. Under the assumption that $\mathbf{U}$
has an extreme value copula, this probability is simply $C\left(b,\ldots,b\right)=b^{\theta}$.
The other probabilities can be very difficult to evaluate in terms
of the original copula, $C$. So we shall try to use this value to
our convenience.

Note that if $b\rightarrow1^{-}$, then 
\[
\frac{\left(1-b^{\alpha}\right)}{\left(1-b\right)^{\alpha}}\rightarrow+\infty
\]
and hence as we approach 1 from below, it is right to assume that
above certain $0<b_{0}<1$, one has $\left(1-b^{\alpha}\right)-\left(1-b\right)^{\alpha}>0$. 

For any $b\in\left[0,1\right]$, one has $\Pr\left(\varsigma_{b}\left(j_{i_{1}}\right),\ldots,\varsigma_{b}\left(j_{i_{K}}\right)\right)\leq\left(1-b\right)$,
regardless of whether $\varsigma\left(j_{i_{k}}\right)=0$ or $\varsigma\left(j_{i_{k}}\right)=1$
for each $j_{i_{k}}$in the index set. Then
\[
1-\left(\Pr\left(\varsigma_{b}\left(j_{i_{1}}\right)=0,\ldots,\varsigma_{b}\left(j_{i_{K}}\right)=0\right)^{\alpha}+\left(2^{J}-1\right)\left(1-b\right)^{\alpha}\right)\leq1-\sum_{j_{i_{1}},\ldots,j_{i_{K}}}\Pr\left(\varsigma_{b}\left(j_{i_{1}}\right),\ldots,\varsigma_{b}\left(j_{i_{K}}\right)\right)^{\alpha}
\]

where the term $\left(2^{J}-1\right)\left(1-b\right)^{\alpha}$ accounts
for the remaining $\left(2^{J}-1\right)$ probability values, apart
from $\Pr\left(\varsigma_{b}\left(j_{i_{1}}\right)=0,\ldots,\varsigma_{b}\left(j_{i_{K}}\right)=0\right)$. 

Thus, using the extreme value copula assumption from equation (\ref{eq:extremal_coeff}),
one can define for $b>b_{0}$ the function 
\begin{multline}
g_{1}\left(b\right):=\frac{1-\left(\Pr\left(\varsigma_{b}\left(j_{i_{1}}\right)=0,\ldots,\varsigma_{b}\left(j_{i_{K}}\right)=0\right)^{\alpha}+\left(2^{J}-1\right)\left(1-b\right)^{\alpha}\right)}{\left(1-b^{\alpha}\right)-\left(1-b\right)^{\alpha}}=\\
\frac{1-\left(b^{\theta\alpha}+\left(2^{J}-1\right)\left(1-b\right)^{\alpha}\right)}{\left(1-b^{\alpha}\right)-\left(1-b\right)^{\alpha}}\leq T_{b}^{\alpha}\label{eq:lower_g}
\end{multline}

Note that we require $b>b_{0}$ to ensure that $\left(1-b^{\alpha}\right)-\left(1-b\right)^{\alpha}>0$.

On the other hand, for $b$ sufficiently large, 
\begin{equation}
g_{2}\left(b\right):=\frac{1-b^{\theta\alpha}}{\left(1-b^{\alpha}\right)-\left(1-b\right)^{\alpha}}=\frac{1-\Pr\left(\varsigma_{b}\left(j_{i_{1}}\right)=0,\ldots,\varsigma_{b}\left(j_{i_{K}}\right)=0\right)^{\alpha}}{\left(1-b^{\alpha}\right)-\left(1-b\right)^{\alpha}}\geq T_{b}^{\alpha}\label{eq:upper_g}
\end{equation}

Then, for any $\alpha>1$, and $b$ such that $b>b_{0}$, one has
\[
g_{1}\left(b\right)\leq T_{b}^{\alpha}\leq g_{2}\left(b\right)
\]

Concerning limits, applying L'H\^opital's rule, 
\[
\lim_{b\rightarrow1}g_{1}\left(b\right)=\lim_{b\rightarrow1}\frac{1-\left(b^{\theta\alpha}+\left(2^{J}-1\right)\left(1-b\right)^{\alpha}\right)}{\left(1-b^{\alpha}\right)-\left(1-b\right)^{\alpha}}=\lim_{b\rightarrow1}\frac{\alpha\left(1-b\right)^{\alpha-1}\left(2^{J}-1\right)-\alpha\theta b^{\alpha\theta-1}}{\alpha\left(1-b\right)^{\text{\ensuremath{\alpha}-1}}-\alpha b^{\alpha-1}}=\theta
\]
and hence $\lim_{b\rightarrow1^{-}}g_{1}\left(b\right)=\theta.$

Similarly, one can see using L'H\^opital's rule that
\[
\lim_{b\rightarrow1^{-}}g_{2}\left(b\right)=\theta
\]

Hence, one must have, for any $\alpha>1$, that 
\begin{equation}
\lim_{b\rightarrow1^{-}}T_{b}^{\alpha}=\theta\label{eq:limit_shown}
\end{equation}
as we wanted to show. 

Now, the Tsallis entropy fulfills $H_{b}^{\alpha}\rightarrow H_{b}$,
as $\alpha\rightarrow1^{+}$. Hence one has 
\begin{equation}
\theta=\lim_{\alpha\rightarrow1^{+}}\left(\lim_{b\rightarrow1^{-}}T_{b}^{\alpha}\right)=\lim_{b\rightarrow1^{-}}S_{b}
\end{equation}
as we wanted to show in this part of the appendix.

\section{Parameters used for example of section \ref{sec:Relation-with-extremal}}

We present here the parameters used for the example at section \ref{sec:Relation-with-extremal}.
These parameters were selected in such a way that for the extremal
coefficient, $\theta$, one has $\theta=2$. 

For the Gumbel copula, the dependence parameter used was $\xi=\frac{\log\left(3\right)}{\log\left(2\right)}\approx1.585$.

For the Student copula, $\nu=2.76733$, and 
\[
\rho=\left(\begin{array}{ccc}
1 & .767 & .759\\
.767 & 1 & .624\\
.759 & .624 & 1
\end{array}\right)
\]

\end{document}